# Google Scholar as a tool for discovering journal articles in library and information science

Dirk Lewandowski

Hamburg University of Applied Sciences, Faculty Design, Media and Information, Department Information, Berliner Tor 5, D – 20249 Hamburg, Germany
E-Mail: dirk.lewandowski@haw-hamburg.de

**Abstract**

Purpose: The purpose of this paper is to measure the coverage of Google Scholar for the Library and Information Science (LIS) journal literature as identified by a list of core LIS journals from a study by Schlögl and Petschnig (2005).

Methods: We checked every article from 35 major LIS journals from the years 2004 to 2006 for availability in Google Scholar (GS). We also collected information on the type of availability—i.e., whether a certain article was available as a PDF for a fee, as a free PDF, or as a preprint.

Results: We found that only some journals are completely indexed by Google Scholar, that the ratio of versions available depends on the type of publisher, and that availability varies a lot from journal to journal. Google Scholar cannot substitute for abstracting and indexing services in that it does not cover the complete literature of the field. However, it can be used in many cases to easily find available full texts of articles already found using another tool.

Originality/value: This study differs from other Google Scholar coverage studies in that it takes into account not only whether an article is indexed in GS at all, but also the type of availability.

**Introduction**

As searching for scientific literature changes (see *College Students' Perceptions of the Libraries and Information Resources: A Report to the OCLC Membership*, 2006; Lewandowski, 2006; Nicholas and Rowlands, 2008), one can see that users often prefer general-purpose Web search engines over specialised search services offered by libraries or database providers. However, it is not only the Web search engines that offer an alternative, but also specialized scientific search engines offered by search engine providers, such as Google, or scientific publishers, such as Elsevier. Libraries now compete with a lot of different search systems such as Scirus, Google Scholar, just to name a few.

From the libraries' perspective, it is important to offer solutions that go beyond indexing the traditional library contents and also include content from the Web (Lewandowski and Mayr, 2006; Lossau, 2004). Regarding user experience, libraries also try approaches that are derived from Web search engines (see Sadeh, 2007). The main problem lies in guiding the user not familiar with high-quality resources and not able to formulate advanced search queries to results suitable for his needs (Lewandowski, 2008).

In this paper, we will study Google's approach to the scientific search engine. Starting in late 2004, Google Scholar (GS) was not the first to open parts of the Academic Invisible Web (Lewandowski and Mayr, 2006), but it was surely the most noticed. There are other scientific search engines as well:

- Elsevier's Scirus,[i] which allows simultaneous searching of documents from the free Web, e.g., open access journals and repositories, along with fee-based content from Elsevier journals (see McKiernan, 2005; Notess, 2005; Scirus White Paper: *How Scirus works*, 2004)). While searching Scirus is free, publisher content is only available for a fee.
- Thomson Scientific Web Plus,[ii] a fee-based scientific search engine from Thomson, intended as a companion to the Science Citation Indices (Web of Science).
- BASE (Bielefeld Academic Search Engine)[iii], which indexes open access repositories as well as the Bielefeld University Library's online catalogue and content from publishers (see Pieper and Summann, 2006). Searching BASE as well as all content is free.

The Academic Invisible Web (AIW), which the scientific search engines try to index, is a subset of the Invisible Web. The term "Invisible Web" (Sherman and Price, 2001) refers to the part of the Web that is not accessible to general search engine crawlers. Bergman (2001) uses the term "Deep Web," which he defines more narrowly as the data available via the Web that is stored in databases. Both terms are used synonymously by now. However, we emphasize the database content, be it publishers' databases of published articles or databases with raw research data.

Lewandowski and Mayr (2006) define the Academic Invisible Web "as consisting of all databases and collections relevant to academia but not searchable by the general Internet search engines" (p. 532). The AIW consists of different kinds of content (Lewandowski and Mayr, 2006, p. 532):
- literature (e.g., articles, dissertations, reports, books);
- data (e.g., survey data); and
- pure online content (e.g., open access documents).

Bergman (2001) estimated the Deep Web as containing 550 billion pages. Lewandowski and Mayr (2006) proved that this number is highly overestimated due to mistakes in using mean averages for the database sizes. They estimate the Academic Invisible Web to contain 20 to 100 billion pages.

This data shows that it is a huge challenge to try to index the complete AIW. Most vendors follow an approach that is much narrower in focus. Google Scholar indexes only scientific books and articles, omitting other contents from the academic Web (such as scientists' homepages or other academic web sites). While for books, only (incomplete) bibliographic information is given, for articles the full text is indexed (where available).

Google Scholar integrates documents from two different "collections": the free Web as well as publishers' and scientific societies' content. Google Scholar takes into account that articles can occur in different instances over the Web. For example, an article could be published first on the author's homepage as a preprint and later in a journal by a commercial publisher. Google Scholar attempts to put these instances together (see fig. 1) and allows selection of the appropriate version.

The leadings question for discussion here is whether GS is suitable as a research tool for finding scientific articles. We chose library and information science as an example. We will present data on the coverage of GS in this discipline, differentiating between the different indexing types (abstract, open access full text, publisher version full text).

This paper is organized as follows: First, we review the literature on Google Scholar, focusing on its use as a research tool and on its coverage of the scientific literature. Then, we present the objectives and the methods of our coverage study, followed by the results. We then discuss these results and come to our conclusions.

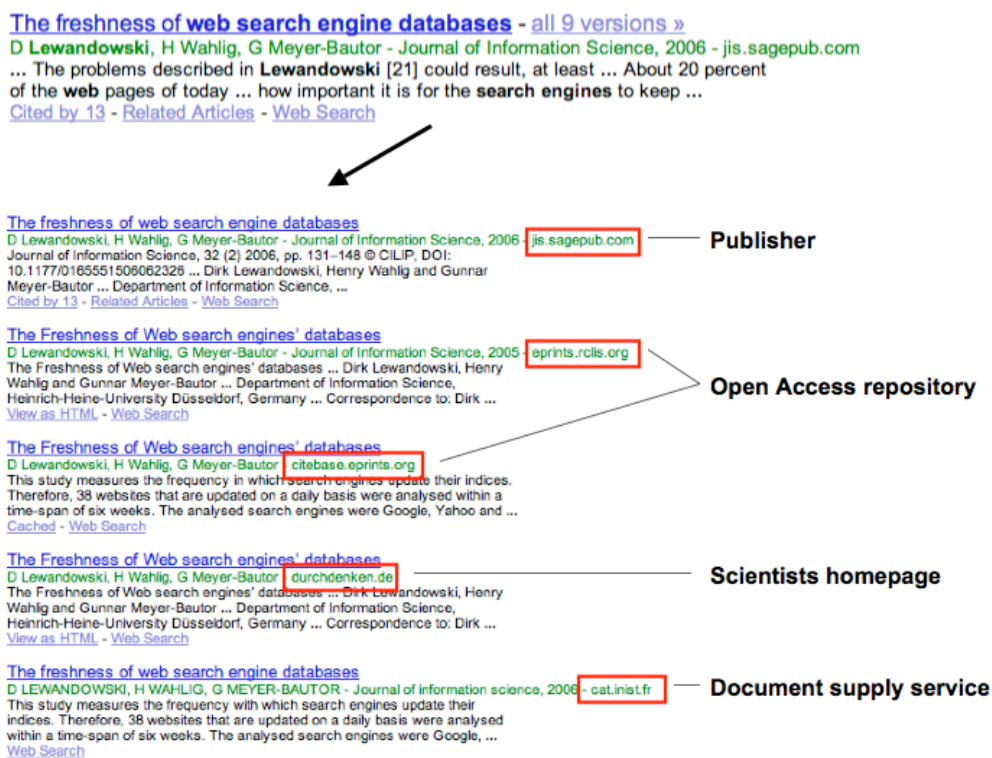

Fig. 1: Different versions of the same article in Google Scholar

**Literature review**

Peter Jacsó (2005 and 2008) gives a good overview of Google Scholar's features and details the problems with this service. While the size of the database increased substantially from 2005 to 2008, Jacsó sees the underlying software as the main problem of Google Scholar. Problems arise not correctly assigned of journal names and authors to the articles. From that arise not only problems with finding articles, but also the conclusion that GS is not useful as a tool for informetric analyses.

Pomerantz (2006) says that libraries should give their users 100 percent availability of information. While Google Scholar could not compete in this respect, this search engine could help to lead users to the library's offerings. Problematic with this approach is that the user often doesn't know or recognize that he indeed uses the library's offerings—i.e., the content the library has licensed for its users. The first as well as subsequent coverage studies made it clear that Google Scholar could not compete when one demands this 100 percent approach to coverage. However, one must say that libraries currently do not give their users this complete coverage through one access point (see Lewandowski, 2006). Using an enormous data set from OIA (Open Archives Initiative) sources, McCown, Nelson, Zubair, and Liu (2006) find that general-purpose Web search engines do not cover a large amount of the scientific literature and are therefore no suitable substitute for specialised databases.

Neuhaus, Neuhaus, Asher, and Wrede (2006) find that GS's coverage has a bias towards the English language and towards science (as opposed to the social sciences and humanities).

Walters (2007) compares Google Scholar to other discipline-specific or multidisciplinary databases based on a set of "155 core articles representing the most important papers on later-life migration published from 1990 to 2000" (p. 1122). The study finds that GS covers 93 percent of the literature, which is better than any of the other databases under investigation. Bar-Ilan (2008) adds to her review of the current informetrics literature (from 2000 onward) an analysis of GS coverage of the 598 articles used for this review. She finds that all but one of these articles could be found using GS. Lewandowski (2007) studies the coverage of GS for articles in the German-language LIS journals. He finds that coverage is comparably low (56 percent), but that for approximately 21 percent of articles, GS can lead the user to the full text.

Meier and Conkling (2008) compare GS's coverage of the engineering literature using Compendex as reference database. They find that the coverage of GS has increased over time, reaching over 90 percent of the Compendex entries for records after 1990. Mayr and Walter (2007) test GS for its coverage of scientific journals based on journal lists from Web of Science (SCI, SSCI, AHCI), the Directory of Open Access Journals, and the SOLIS[iv] database. They find that the coverage of journals (i.e., at least one article for a certain journal was found in GS) varies between databases (from 68 to 88 percent). From the studies reported, we have found that while GS's coverage for certain disciplines varies, the overall coverage is quite good.

Callicott and Vaughn (2006) try to compare Google Scholar to a library OPAC, EBSCO's Academic Search Premier database and a discipline-specific abstracting and indexing service. Unfortunately, they use only five tasks, which makes the results only exploratory. However, the test does show that although Google Scholar does not perform well, results vary from one task to the other. GS is, by far, not a substitute for other abstracting and indexing services. Another interesting approach is to ask which search service works best to advance users' knowledge on a certain topic. Machill, Beiler, and Neumann (2007) evaluate three services—Google Web, Google Scholar, and the local library—in this regard. Users are presented with one of two research questions (from communications). They have 15 minutes time to find and evaluate results with one of the search services. Given this short time-span, it comes as no surprise that Google Web performs best. Web search results usually offer easier-to-read, overview articles in opposition to the highly specific scientific texts that are usually found in A&I or full-text databases. However, the approach to measuring the increase of knowledge seems to be promising and should be kept in mind for future studies.

While the comparison of GS and other (scientific) search engines is an interesting topic, the research on this question is still in its infancy. Further research is needed to find out how a good combination of such services with other more traditional sources could be achieved.

**Objectives and methods**

With this study, we wanted not only to find out whether the LIS journal literature is indexed in GS, but also in which form. Therefore, we distinguished between different forms of the articles—i.e., abstract, preprint full text, free PDF publisher version, and fee publisher version.

The methods of our study followed those in our study on German-language LIS journals in Google Scholar (Lewandowski, 2007), where the last three volumes of the major German LIS journals were used as the dataset. Each article was checked in GS and type of coverage was taken down. We found that the coverage ration for individual journals differs widely and that on average, 56 percent of articles could be found using GS. The ratio of articles available as preprints was very low (2.5 percent), which indicates that German-language LIS authors do not publish their preprints in repositories or on their homepages, not that GS is unable to index these preprints. Spot checks indicate that GS is well able to find and index these preprints. A surprising result of that study was the high number of articles available as a free PDF (18.3 percent), whether freely available from the publisher (sometimes available after a certain length of time from publication) or available from the authors' homepages (whether permitted by the publisher or not). In total, approximately 21 percent of all articles were available in full text in one or the other form.

*Selection of journals*

To generate our data set, we used the list of core LIS journals from Schlögl and Petschnig (2005), which is a selection from the LIS category in the Journal Citation Reports, supplemented with the main German-language journals. We omitted the latter from our current study, as we already had analyzed them in our past study. In total, our data set for the current study consisted of 35 journals. The journal list can be found in the appendix to this article.

In addition to selecting articles from the journals (see below), we researched the open access policy of the journal from the publishers' Web sites and the Sherpa Romeo database,[v] respectively.

*Selection of articles*

For our study, we wanted to use articles only—i.e., we wanted to omit news items and editorials that might be relevant only for a certain length of time. We assume that authors will make available only their more substantial work in full text on their homepages or in institutional repositories. In addition, we wanted to restrict the articles to those of scientific interest.

Where articles are not labelled exactly enough (i.e., whether individual articles were research articles, news items, etc.) in the table of contents, we collected the articles from the Web of Science database (for the journals included in that database). For the study, we used all articles published in the years ranging from 2004 to 2006.

Data collection led us to 3,580 English language articles, which is the complete set of all articles from the journals selected which were published from 2004 to 2006. For comparisons, we were also able to use 791 German language articles from our previous study (Lewandowski 2007).

*Classification of articles*

We classified every article in the named journals according to indexing type:
- Bibliographic data: These are records without a link to any kind of full text—i.e., books cited in an article's index by Google Scholar or citations from indexed articles where no further information could be found.
- Preprint: A preprint here refers to any author version of an article that is not in the publisher's citable format. Most publishers allow for such versions to be published on authors' homepages or in institutional repositories. We do not differentiate between preprint and postprint versions of an article.
- Free PDF: This means that a PDF file in the original layout of the journal is available without cost. On the one hand, these can be articles provided by the publisher for free (for marketing purposes or because of a publisher's policy to provide free access to its articles after a certain length of time[vi]). On the other hand, some authors simply ignore their copyright agreement with their publisher and make available the publisher's version of their articles on their homepages.

- PDF available for a fee: This means that the publisher's version is available through Google Scholar. If this is the case, this version is shown first in the list of instances of a certain article. When users click on the result in the results list without selecting the instances of an article, the publisher's version is shown. However, access depends on an institutional subscription. If the user's institution is subscribed to the journal selected, the user is taken directly to the full text. If not, the publisher's notice for restricted access is shown.

*Data collection*

We searched Google Scholar for all articles from our journal set that were published between 2004 and 2006. We collected data on the availability of the articles according to our article classification. Data were collected in September and October 2007. We were not able to collect data within a shorter period as GS restricts the number of queries from an individual IP address within a certain time-span. However, we do not think that the longer time-span affected our results. We monitored GS for index updates, but we did not observe any in the time of our investigation. Therefore, the results would have been the same, even if we had been able to collect the data in just one day.

**Results**

This section shows the main results from our study. Complete data for all journals can be found in the appendix.

*Availability of articles*

First, it is of interest to what degree GS is able to give a complete picture of the field—i.e., what ratio of LIS articles is covered by GS. Regardless of journal, GS found 3,279 articles of our total set (91.6 percent). While this is an impressive amount, it also clearly shows that GS is no substitute for abstracting and indexing databases (like Web of Science or field-specific databases such as LISA) that cover 100 percent of research articles from the journals indexed.

Looking at the different journals, we found that there is a wide distribution in coverage. Eight journals (of 35 in total) are covered 100 percent. These are *Canadian Journal of Information Science, College and Research Libraries, Information Technology & Libraries, Journal of Information Ethics, Journal of Information Science, Journal of Librarianship and Information Science, Library Quarterly,* and *Online Information Review*.

It is evident that some of the major LIS journals are not covered in entirety. For most journals, coverage rate lies well over 90 percent (22 journals), but for four journals, the coverage is below 70 percent. Among these is ARIST, one of the major LIS periodicals.

Looking at the availability of preprint versions of articles, we found that only a low number of articles is available in this form. Only 353 articles of our dataset are available as preprints (9.9 percent). Random examination indicates that this is not Google Scholar's fault but that of the authors. Only a limited number of authors make their articles available through institutional or discipline-specific repositories[vii] or on their homepages.

The highest ratio of preprints can be found for Information Processing & Management (28.1%), Program (23.4%), and Journal of the American Society for Information Science & Technology (20.4%). With 14 titles, the largest group of journals has a ratio of preprints from 10 to 20 percent.

Eight journals have a ratio from five to ten percent, but a group of 10 journals have less than five percent of their articles available as preprints.

The numbers for the preprints are somehow disappointing as there are high hopes for Open Access and the willingness of authors to make their work available through OA. Especially for the LIS profession with its many OA promoters, the numbers seem to be very low.

In the case of free PDF versions of an article (see above for definition), only one journal publisher provides free PDFs of older articles, i.e. the American Library Association with *College and Research Libraries*. Articles in this journal are made freely available six months after publication.

All other free PDFs consist of publisher files made available by the authors with or without consent from the publisher. While we were not able to determine whether the publisher allowed any given individual article to be put on the Web, we assume that the majority of these articles were made available without the consent of the publisher. To our knowledge, none of the commercial publishers allows for putting such PDF files on the Web.

In total, 375 articles (10.4 percent) were made available as a free PDF. Seventy-three of these were from *College and Research Libraries*, which unsurprisingly has the largest ratio of articles available in this form (81.1 percent). It is surprising that the number of freely available publisher PDF files is larger than the number of preprint files (353 articles or 9.9 percent).

Excluding C&RL from our dataset, we continued only with the articles made available by the authors themselves. The journals with the highest ratio of free PDF files are *Libri* (41.8%), *Information Technology & Libraries* (33,3%) and *Library Hi Tech* (22.5%).

For nine journals, between 10 and 20 percent of articles are available in this form, while the largest group of journals (22) have less than ten percent of their articles available. For five of these journals, we did not find a single article available as a free PDF.

Making publisher PDF files available on the Web requires that the authors have access to these files or scan the paper documents, respectively. While all the major publishers obtain Web archives where authors can download the files and make them available on their personal homepages, smaller publishers, in particular, do not obtain such archives. Publishers have different policies with regard to providing authors with PDF files of their articles. In our experience, publishers do not routinely send the authors the final version of the articles as a PDF.

*Availability of the articles with respect to the publisher*

The journals analyzed in this study were published by a total of 16 publishers. We queried the Romeo Sherpa database for publisher information, as this database does not only list the publishers but also the Open Access policies for each individual publisher and journal. The publishers were either classified as *green* ("can archive pre-print and post-print") or *yellow* ("can archive pre-print (i.e., pre-refereeing)").

The results are shown in table 3. While there are no significant differences in the availability of free PDFs or fee PDFs, whether the publisher has a green or a yellow OA policy, in the percentage of preprints, there is a slightly lower rate of preprints.

Regarding the publishers where no information on their OA policy is available, we found that the percentage of available articles is much lower in all categories.

Table 1: Availability of articles according to open access policy of publisher

| OA policy | Number of articles | Free PDF | Preprint | PDF (fee) |
|---|---|---|---|---|
| Green | 2246 | 13.1% | 12.2% | 75.6% |
| Yellow | 631 | 12.0% | 9.8% | 73.4% |
| No information available | 703 | 0.6% | 2.4% | 0.6% |

Additionally, we classified the publishers into commercial operations, university publishers, and societies. One could assume that societies or university publishers encourage their authors to self-archive their articles or to make them available on their homepages, while commercial publishers would do more to make their articles show up in search engines like Google or Google Scholar in the form of fee-based PDFs. Indeed, we can see from the results (table 2) that about two thirds of articles from commercial publishers are available in this form. However, 62.2 percent of articles from university presses are available for a fee, too. None of the articles from society publishers is available for a fee. However, articles from this group of publishers show the largest percentage of articles available as a free PDF. This could come from the lack of commercial interest of the societies. Availability of free PDFs from commercial publishers and university presses is nearly the same (at around eight percent).

Table 2: Availability of articles according to commercial intent of publisher

| Publisher intent | Number of articles | Free PDF | Preprint | PDF (fee) |
|---|---|---|---|---|
| Commercial | 3063 | 8.2% | 10.4% | 67.2% |
| Society | 300 | 35.3% | 5.7% | 0% |
| University | 217 | 8.3% | 7.4% | 62.2% |

Looking at the results for the individual publishers (table 3), we found that articles published with one of the large commercial publishers have a very high chance to be indexed by Google Scholar. More than 90 percent of articles from Emerald, Wiley, Springer, Sage, and Taylor & Francis are indexed in Google Scholar. The exception here is Elsevier. Only 71.1 percent of articles from this publisher were found using Google Scholar. Smaller publishers and societies seem not to have powerful archives that are available to search engines like GS; and, therefore, their articles are not indexed to a similar degree. For some publishers, there are no articles offered for a fee.

Regarding the availability of preprints, we also found differences between the publishers. The largest ratio of articles in this form is from Wiley (20.4 percent), Taylor & Francis (18.9 percent) and Ergon (18.4 percent). However, these publishers publish only one journal each, so the total numbers are low.

Looking at the two biggest players in the LIS field (Emerald and Elsevier, according to articles published), we found that the ratio of preprints lies between 11.6 and 13.2 percent.

When looking at the availability of free PDFs, we found that K.G. Saur and ALA lead with more than 40 percent of articles available in this form. This clearly results from these publishers' policy to make articles freely available a certain length of time after publication. However, articles from other publications are available for free in their original PDF form. Sometimes, this could result from publishers' making available certain articles (or journal issues) for free. However, the larger number of articles available in this form, by far, results from authors' making available the original PDFs on their homepages, whether the publisher has given permission or not.

Table 3: Availability of articles according to publisher

| Publisher intent | Number of articles | Free PDF | Preprint | PDF (fee) |
| --- | --- | --- | --- | --- |
| Alise Association for Library and Information | 48 | 6.3% | 6.3% | 0.0% |
| American Library Association | 252 | 40.9% | 5.6% | 0.0% |
| Elsevier | 696 | 7.2% | 13.2% | 71.1% |
| Emerald | 819 | 10.3% | 11.6% | 94.4% |
| Ergon | 38 | 0.0% | 18.4% | 0.0% |
| Information Today | 249 | 0.4% | 2.8% | 0.4% |
| John Wiley and Sons | 225 | 16.0% | 20.4% | 99.1% |
| K.G. Saur | 67 | 41.8% | 4.5% | 0.0% |
| Mcfarland & Company, Inc. | 33 | 0.0% | 3.0% | 100.0% |
| Reed Business Information (division of Reed Elsevier) | 379 | 0.0% | 0.5% | 0.3% |
| SAGE Publications | 257 | 7.8% | 10.1% | 93.0% |
| Springer-Verlag | 210 | 10.0% | 11.4% | 97.6% |
| Taylor & Francis | 90 | 11.1% | 18.9% | 98.9% |
| University of Chicago Press | 61 | 9.8% | 13.1% | 78.7% |
| University of Illinois Press | 86 | 9.3% | 5.8% | 29.1% |
| University of Toronto Press | 70 | 5.7% | 4.3% | 88.6% |

**Discussion and Conclusion**

The results show that for only eight journals, Google Scholar is able to index 100 percent of the articles, while for the other journals, no complete coverage is given. However, for most journals, the coverage ratio is well over 95 percent. This leads to the conclusion that while lots of LIS articles can be found using Google Scholar, it is not a substitute for abstracting and indexing databases that give complete coverage of all articles of the journals selected for indexing. Also, it must be kept in mind that the sample used in this study is just a selection of LIS journals. Some A&I databases (like Library & Information Science Abstracts (LISA)) offer a far wider coverage of LIS journals. From our study, we cannot say whether the coverage of other journals (from which some would not be considered as core LIS journals) in Google Scholar is similar to that of the core journals used in our study.

However, only relatively slight improvements could lead to a complete coverage of the journals investigated in Google Scholar. Then the advantages of A&I services for the given journal set would remain only in the better search tools if one would settle for searching only the core journals of the field.

One main advantage of GS is the direct availability of articles, independent of where they are deposited. When searching for an article, one can often choose between the publisher's version and a preprint. In many cases, one can find at least one free version of an article and in many cases, even the original (publisher) PDF. This leads to the question whether publishers should tolerate such a use of their articles. From an author's point of view, it is most important that their articles get read. They aim for the widest distribution, whether through the publisher or through making a copy available on the Web. It is interesting to see that for many journals, the ratio of free PDFs is higher than the ratio of preprints. Aside from cases in which free PDFs are made available by the publisher itself, authors may either think that they are allowed to publish the original PDF, or they may not care about what is stated in the publishing contract. As the publishers' main capital is their authors, it is difficult to ask authors who publish the original PDFs on their Web sites to remove them. Publishers may be right to do this from a legal point of view, but in doing so, they would risk annoying their authors (and in consequence, lose them).

The low ratio of articles available as preprints may result either from the authors' unwillingness to deposit them or from the inability of Google Scholar to index them. As we can see from sample searches, the latter seems not to be the case. This leads to the conclusion that while many researchers and practitioners in the LIS field are advocates for Open Access, when it comes to helping Open Access through depositing preprints, their support ends.

Comparing the results of the present study with our study on German-language LIS articles (Lewandowski, 2007), we can say that a much higher ratio of English-language LIS articles is available in Google Scholar. However, German journals are published mainly by small publishers, and for some no online version is available. This explains the smaller coverage rate sufficiently. The results do not indicate a language or country bias towards English-language articles.

The approach taken in this paper is new in that it widens the coverage studies conducted so far to include the different indexing types. With this approach, we were able to show that while GS may not be the best tool to find the LIS literature, it can greatly help to achieve the full text of the desired publications.

**References**

Bar-Ilan, J. (2008), "Informetrics at the beginning of the 21st century: A review", *Journal of Informetrics,* Vol. 2 No. 1, pp. 1-52.
Bergman, M. K. (2001), "The deep Web: Surfacing hidden value", *Journal of Electronic Publishing,* Vol. 7 No. 1.
Callicott, B., & Vaughn, D. (2006), "Google scholar vs. library scholar: Testing the performance of schoogle", *Internet Reference Services Quarterly,* Vol. 10 No. 3/4, pp. 71-88.
"College students' perceptions of the libraries and information resources: A report to the OCLC membership"(2006), OCLC, Dublin.
Jacsó, P. (2005), "Google Scholar: The pros and cons", *Online Information Review,* Vol. 29 No. 2, pp. 208-214.
Jacsó, P. (2008), "Google Scholar revisited", *Online Information Review,* Vol. 32 No. 1, pp. 102-114.
Lewandowski, D. (2006), "Suchmaschinen als Konkurrenten der Bibliothekskataloge: Wie Bibliotheken ihre Angebote durch Suchmaschinentechnologie attraktiver und durch Öffnung für die allgemeinen Suchmaschinen populärer machen können", *Zeitschrift für Bibliothekswesen und Bibliographie,* Vol. 53 No. 2, pp. 71-78.
Lewandowski, D. (2007), "Nachweis deutschsprachiger bibliotheks- und informationswissenschaftlicher Aufsätze in Google Scholar", *Information Wissenschaft und Praxis,* Vol. 58 No. 3, pp. 165-168.
Lewandowski, D. (2008), "Search engine user behaviour: How can users be guided to quality content?", *Information Services & Use,* Vol. 28, pp. 261-268
Lewandowski, D. and Mayr, P. (2006), "Exploring the academic invisible web", *Library Hi Tech,* Vol. 24 No. 4, pp. 529-539.
Lossau, N. (2004), "Search engine technology and digital libraries: Libraries need to discover the academic Internet", *D-Lib Magazine,* Vol. 10 No. 6. http://www.dlib.org/dlib/june04/lossau/06lossau.html

Machill, M., Beiler, M. and Neumann, U. (2007), "Leistungsfähigkeit von wissenschaftlichen Suchmaschinen. Ein Experiment am Beispiel von Google Scholar", in Machill, M. and Beiler, M. (Eds.), *Die Macht der Suchmaschinen / The power of search engines*, Herbert von Halem, Köln, pp. 327-347.

Mayr, P. and Walter, A.-K. (2007), "An exploratory study of Google Scholar", *Online Information Review,* Vol. 31 No. 6, pp. 814-830

McCown, F., Nelson, M.L., Zubair, M. and Liu, X. (2006), "Search engine coverage of the OAI-PMH corpus", *IEEE Computer,* Vol. 10 No. 2, pp. 66-73.

McKiernan, G. (2005), "E-profile: Scirus: For scientific information only", *Library Hi Tech News,* Vol. 22 No. 3, pp. 18-25.

Meier, J.J. and Conkling, T.W. (2008), "Google Scholar's coverage of the engineering literature: An empirical study", *Journal of Academic Librarianship,* Vol. 34 No. 3, pp. 196-201.

Neuhaus, C., Neuhaus, E., Asher, A. and Wrede, C. (2006), "The depth and breadth of Google Scholar: An empirical study", *Portal-Libraries and the Academy,* Vol. 6 No. 2, pp. 127-141.

Nicholas, D. and Rowlands, I. (Eds.), (2008), *Digital consumers: Re-shaping the information professions*, Facet, London.

Notess, G.R. (2005), "Scholarly Web searching: Google Scholar and Scirus", *Online,* Vol. 29 No. 4, pp. 39-41.

Pieper, D. and Summann, F. (2006), "Bielefeld Academic Search Engine (BASE): An end-user oriented institutional repository search service", *Library Hi Tech,* Vol. 24 No. 4, pp. 614-619.

Pomerantz, J. (2006), "Google Scholar and 100 percent availability of information", *Information Technology and Libraries,* Vol. 25 No. 1, pp. 52-56.

Sadeh, T. (2007), "Time for a change: new approaches for a new generation of library users", *New Library World,* Vol. 108 No. 7/8, pp. 307-316.

Schlögl, C. and Petschnig, W. (2005), "Library and information science journals: An editor survey", *Library Collections, Acquisitions, & Technical Services,* Vol. 29 No. 1, pp. 4-32.

"Scirus White Paper: How Scirus works" (2004)*,* available at: http://www.scirus.com/press/pdf/WhitePaper_Scirus.pdf (accessed 8 March 2006).

Sherman, C. and Price, G. (2001), *The invisible Web: Uncovering information sources search engines can't see,* Information Today, Medford, NJ.

Stock, W.G. and Schlögl, C. (2004), "Impact and relevance of LIS journals: A scientometric analysis of international and German-language LIS journals - Citation analysis versus reader survey", *Journal of the American Society for Information Science and Technology,* Vol. 55 No. 13, pp. 1155-1168.

Walters, W.H. (2007), "Google Scholar coverage of a multidisciplinary field", *Information Processing & Management,* Vol. 43 No. 4, pp. 1121-1132.

**Appendix**

| Journal | Number of articles | Found in Google Scholar | Available as a preprint (percent) | Available as a free PDF (percent) | Available as a PDF for a fee (percent) |
|---|---|---|---|---|---|
| Annual Review of Information Science and Technology | 44 | 61.4 | 6.8 | 2.3 | 2.3 |
| ASLIB Proceedings | 113 | 99.1 | 7.1 | 3.5 | 97.3 |
| Canadian Journal of Information Science | 25 | 100.0 | 12.0 | 0.0 | 0.0 |
| College and Research Libraries | 90 | 100.0 | 1.1 | 81.1 | 0.0 |
| E-Content | 109 | 66.1 | 0.0 | 0.0 | 0.0 |
| Electronic Library. The | 139 | 97.8 | 11.5 | 2.9 | 97.1 |
| Government Information Quarterly | 89 | 94.4 | 7.9 | 4.5 | 94.4 |
| Information Processing & Management | 167 | 98.8 | 28.1 | 16.2 | 96.4 |
| Information Society | 90 | 98.9 | 18.9 | 11.1 | 98.9 |
| Information Technology & Libraries | 69 | 100.0 | 11.6 | 33.3 | 0.0 |
| Interlending & Document Supply | 101 | 96.0 | 5.9 | 8.9 | 96.0 |
| International Journal of Information Management | 111 | 98.2 | 9.0 | 6.3 | 0.0 |
| Journal of Academic Librarianship | 167 | 98.8 | 4.8 | 3.6 | 98.2 |
| Journal of Documentation | 110 | 98.2 | 10.9 | 16.4 | 98.2 |
| Journal of Education for Library and Information Science | 48 | 95.8 | 6.3 | 6.3 | 0.0 |
| Journal of Information Ethics | 33 | 100.0 | 3.0 | 0.0 | 100.0 |
| Journal of Information Science | 125 | 100.0 | 15.2 | 10.4 | 100.0 |
| Journal of Librarianship and Information Science | 49 | 100.0 | 10.2 | 12.2 | 87.8 |
| Journal of Scholarly Publishing | 70 | 98.6 | 4.3 | 5.7 | 88.6 |
| Journal of the American Society for Information Science and Technology | 225 | 99.6 | 20.4 | 16.0 | 99.1 |
| Knowledge Organization | 38 | 94.7 | 18.4 | 0.0 | 0.0 |
| Library & Information Science Research | 87 | 96.6 | 12.6 | 5.7 | 96.6 |
| Library Collections. Acquisitions and Technical Services | 75 | 98.7 | 12.0 | 1.3 | 2.7 |
| Library Hi Tech | 129 | 99.2 | 14.7 | 22.5 | 98.4 |
| Library Journal | 379 | 56.2 | 0.5 | 0.0 | 0.3 |
| Library Quarterly | 61 | 100.0 | 13.1 | 9.8 | 78.7 |
| Library Resources & Technical Services | 49 | 93.9 | 4.1 | 10.2 | 0.0 |
| Library Trends | 86 | 94.2 | 5.8 | 9.3 | 29.1 |
| Libri | 67 | 98.5 | 4.5 | 41.8 | 0.0 |
| Online | 96 | 84.4 | 4.2 | 1.0 | 0.0 |
| Online Information Review | 125 | 100.0 | 10.4 | 15.2 | 99.2 |
| Program | 77 | 96.1 | 23.4 | 1.3 | 93.5 |
| Reference & User Services | 44 | 61.4 | 6.8 | 4.5 | 0.0 |
| Quarterly | | | | | |
| Scientometrics | 210 | 99.0 | 11.4 | 10.0 | 97.6 |
| Social Science Information | 83 | 97.6 | 2.4 | 1.2 | 85.5 |

---

[i] http://www.scirus.com

[ii] http://scientific.thomsonwebplus.com

[iii] http://base.ub.uni-bielefeld.de/

[iv] SOLIS is a mainly German language literature database in the social sciences.

[v] http://www.sherpa.ac.uk/romeo.php

[vi] There were no such articles in our current study. However, some articles from the German-language LIS journals (see below) are available for free from the publishing bodies.

[vii] Such as E-LIS (rclis.eprints.org) or D-LIST (dlist.sir.arizona.edu). At the time of writing, E-LIS contained 8800 articles and D-LIST 1380, respectively. Both archives host not only journal articles, but also conference papers, presentations, and other manuscripts.